\documentclass[12pt]{article}

\usepackage{amssymb, amsmath, amsthm}
\usepackage{graphicx,subfig,tikz}
\usepackage{cite}
\usepackage[left=2cm,right=2cm,top=2cm,bottom=2cm]{geometry}

\usepackage[colorlinks=true,linktocpage=true,allcolors=blue]{hyperref}

\def\be{\begin{equation}}
\def\ee{\end{equation}}
\def\bea{\begin{eqnarray}}
\def\eea{\end{eqnarray}}

\newcommand{\td}{\text{d}}

\title{ \bf{On the nonexistence of a vacuum black lens}}

\author{James Lucietti\footnote{j.lucietti@ed.ac.uk}\,  and Fred Tomlinson\footnote{f.tomlinson@ed.ac.uk}
\\ \\ \small \sl School of Mathematics and Maxwell Institute for Mathematical Sciences, \\ \small \sl    University of Edinburgh, King's Buildings, Edinburgh, EH9 3JZ, UK }

\date{}

\begin{document}

%\begin{titlepage}
\maketitle

\begin{picture}(0,0)(0,0)
\put(350, 240){}
\put(350, 225){}
\end{picture}

%\vskip1.5cm
%Abstract
\begin{abstract}
We demonstrate that five-dimensional, asymptotically flat, stationary and biaxisymmetric, vacuum black holes with lens space $L(n,1)$ topology, possessing the simplest rod structure, do not exist. In particular, we show that the general solution on the axes and horizon, which we recently constructed by exploiting the integrability of this system, must suffer from a conical singularity on the inner axis component. We give a proof of this for two distinct singly spinning configurations and numerical evidence for the generic doubly spinning solution.
\end{abstract}

\newpage 
%\tableofcontents

\newpage 
\section{Introduction}

Black holes in General Relativity (GR) must have spherical horizon topology.  This fundamental result follows from Hawking's horizon topology theorem, which  shows that asymptotically flat, stationary, black hole spacetimes obeying the dominant energy condition have an event horizon with cross-sections of $S^2$ topology~\cite{Hawking:1971vc}.  A striking property of higher-dimensional GR is that the topology of black holes is not so tightly constrained.  This was first revealed by the discovery of a five-dimensional, asymptotically flat, vacuum spacetime containing a black hole with $S^1\times S^2$ topology, known as a black ring~\cite{Emparan:2001wn}.  Together with the $S^3$ topology Myers-Perry black hole~\cite{Myers:1986un}, this also demonstrates that the black hole uniqueness theorem does not hold in higher dimensions in any simple form.  This motivated the classification of higher-dimensional black holes, which remains largely an open problem~\cite{Emparan:2008eg, Hollands:2012xy}.

We will consider this classification problem in the most basic higher-dimensional set-up: five-dimensional, asymptotically flat, stationary vacuum spacetimes, that admit two commuting axial Killing fields (biaxial symmetry). This is a well-studied class~\cite{Emparan:2001wk,Harmark:2004rm,Hollands:2007aj, Hollands:2008fm} (see also review~\cite{Hollands:2012xy}), which includes both the black ring and the Myers-Perry black hole. The metric can be written in terms of Weyl-Papapetrou coordinates,
\be
\mathbf{g}= g_{AB}\td x^A \td x^B+ e^{2\nu} (\td \rho^2+ \td z^2)  \; , \label{WP}
\ee
where $A, B \in \{ 0,1, 2 \}$, $\det g = - \rho^2$ and $k=\partial_0,\ m_i= \partial_i,\ i \in \{ 1,2 \}$ are the stationary and biaxial Killing fields respectively.  It has been shown this is a global chart away from the axes and horizons and the orbit space for such spacetimes can be identified with a half-plane $\{ (\rho, z) \; | \; \rho>0 \}$. The boundary of the orbit space (the $z$-axis) divides into intervals, called rods, which correspond  to either components of the axis where a periodic linear combination of the biaxial Killing fields vanishes (called the axis rod vectors), or to components of the event horizon. The collection of this boundary data is known as the {\it rod structure} and it encodes both the horizon and spacetime topology.  The only possible (connected) horizon topologies are $S^3, S^2\times S^1$ and lens spaces $L(p,q)$.  

A uniqueness theorem for this class of spacetimes has been known for some time, which reveals that the angular momentum and rod structure is sufficient to uniquely characterise black hole solutions~\cite{Hollands:2007aj, Hollands:2008fm}.   The known solutions realise the simplest rod structures compatible with their topology. More recently an existence theorem has been proven which guarantees existence of solutions up to potential conical singularities on the inner axis rods~\cite{Khuri:2017xsc}.  Despite these advances, the existence question for {\it regular} solutions has remained open, i.e., for what angular momenta and rod structures do regular black hole solutions actually exist?  This is a nontrivial question since, unlike the case of four dimensions~\cite{Neugebauer:2011qb}, one can have regular solutions with nontrivial rod structure corresponding to multi-black holes~\cite{Elvang:2007rd, Iguchi:2007is, Izumi:2007qx, Elvang:2007hs}. Furthermore, single black hole solutions with arbitrarily complicated rod structures can also exist in principle (indeed, this is known to be the case for supersymmetric solutions~\cite{Breunholder:2017ubu}).

Recently, we derived the general solution on the axes and horizons for any given rod structure~\cite{Lucietti:2020ltw}. The method uses the integrability of the Einstein equations for this symmetry class of metrics. That is, we integrated the Belinski-Zakharov linear system~\cite{Belinsky:1971nt} along the axes and horizons and around infinity, which then allows one to fully determine the metric data on the axes and horizons. In Weyl-Papapetrou coordinates the solutions are rational functions of $z$, which depend explicitly on several geometrically defined moduli: the rod lengths,  axis rod vectors, horizon angular momenta and angular velocities, and certain gravitational fluxes. These parameters must obey certain nonlinear algebraic equations -- the moduli space equations -- which are difficult to solve explicitly in general.   Generically, the solutions will possess conical singularities on the axes, and imposing their absence leads to further constraints on the parameters.  For the simplest rod structures corresponding to the Myers-Perry black hole and the black ring, one can solve the moduli space equations and obtain uniqueness proofs for the known solutions~\cite{Myers:1986un, Emparan:2001wn, Pomeransky:2006bd}.

In this note we will consider the most basic open question in this context:  do regular black holes with lens space topology exist? A number of attempts at constructing such black lens solutions have resulted in singular spacetimes, the mildest being a conical singularity on the inner axis~\cite{Evslin:2008gx, Chen:2008fa, Tomizawa:2019acu}. However, it has remained unclear whether these past works represent the most general solutions with their given rod structures, thereby leaving the question of existence unanswered. To address this question, we will analyse regularity of the general solution on the axes and horizon recently found in~\cite{Lucietti:2020ltw} for the simplest rod structure corresponding to a $L(n,1)$ black lens  (see Figure \ref{fig:BL}). As we will explain below, by a mix of analytic and numerical analysis, we will show that regular black lenses of this type do not in fact exist. In particular, they must possess a conical singularity on the inner axis rod.

This note is organised as follows. In section 2 we recall the general solution found in~\cite{Lucietti:2020ltw} specialised to the black lens rod structure and derive the conditions for the absence of conical singularities on the axis. In section 3 we solve the moduli space equations analytically for two distinct singly spinning cases and prove that such solutions must be conically singular; we also show that the moduli space at large $n$ reduces to that of the Myers-Perry black hole. In section 4 we numerically solve the moduli space equations for the generic doubly spinning black lens  and present numerical evidence  that these solutions are always conically singular on the inner axis. We close with a discussion of our results.

\section{Black lens solution}

We  consider the rod structure for the simplest $L(n,1)$ black lens as in Figure \ref{fig:BL} below.\footnote{For $n\neq 0$ we define $L(n,1)= S^3/ \sim$ where $\sim$ is given by $w \sim e^{2 \pi i/n} w$
%,$w \in S^3 = \{ |w_1|^2+|w_2|^2 \} \subset  \mathbb{C}^2$. 
and $w \in \{(w_1,w_2) \in \mathbb{C}^2 : |w_1|^2+|w_2|^2 = 1\} \cong S^3$. For $n=0$ we define $L(0,1)= S^1\times S^2$.
}
\begin{figure}[h!]
\centering
\subfloat{
\begin{tikzpicture}[scale=1.5, every node/.style={scale=1}]
\draw[very thick](-4,0)--(-0.9,0)node[black,left=2.1cm,above=.2cm]{$(0,1)$};
\draw[thick,dashed](-.7,0)--(.7,0)node[black,left=1.0cm,above=.2cm]{$H$};
\draw[very thick](0.9,0)--(2.0,0)node[black,left=.8cm,above=.2cm]{$(n,1)$};
\draw[very thick](2.2,0)--(4,0)node[black,left=1.4cm,above=.2cm]{$(1,0)$};
\draw[fill=white] (-.8,0) circle [radius=.1] node[black,font=\large,below=.1cm]{};
\draw[fill=white] (.8,0) circle [radius=.1] node[black,font=\large,below=.1cm]{};
\draw[fill=black] (2.1,0) circle [radius=.1] node[black,font=\large,below=.1cm]{};
\end{tikzpicture}}
\caption{Rod structure for a black lens with $L(n,1)$ horizon topology.}
\label{fig:BL}
\end{figure}
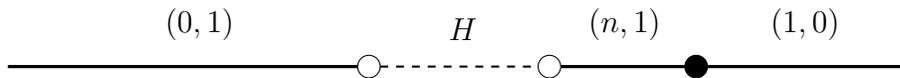
 Thus we have four rods: $I_L=(-\infty, z_1)$, $I_H=(z_1, z_2)$, $I_D=(z_2, z_3)$ and $I_R=(z_3, \infty)$, with $z_1<z_2<z_3$, where  $I_H$ is a horizon rod and $I_L, I_D, I_R$ are axis rods. We choose the semi-infinite axis rod vectors to be $v_L=(0,1)$ and $v_R=(1,0)$ and the finite axis rod vector to be $v_D= (n,1)$ where $n\in \mathbb{Z}$, relative to the basis of biaxial Killing fields $(m_1, m_2)$, both normalised with $2\pi$-period.  Indeed, up to irrelevant  discrete choices this is the most general rod structure with one horizon and one finite axis rod, that is compatible with asymptotic flatness and  obeys the admissibility condition $\det ( v_D, v_R) =\pm 1$ (this latter condition ensures the absence of orbifold singularities at $z=z_3$)~\cite{Hollands:2007aj}. The topology of the horizon is the lens space $L(n,1)$  and the finite axis rod $I_D$ lifts to a noncontractible 2-disc in the spacetime. For $n=0$ this is the rod structure for the black ring solution.  For $n=\pm 1$ the horizon topology $L(1,1)\cong S^3$ is spherical, although the rod structure is distinct to that of the Myers-Perry black hole. 

\subsection{General solution on axes and horizon}
\label{sec:gensol}

We now recall the general solution on the axes and horizon constructed in~\cite{Lucietti:2020ltw}.
 The metric  is given by (\ref{WP}) where the matrix of inner products of Killing fields $g(z):= g|_{\rho=0}$ on each rod $I_a$\footnote{In some of the formulas we label the rods $I_a$ numerically so $a=1,2,3,4$ corresponds to $L,H,D,R$ respectively.}, relative the standard basis $E_A :=(k, m_1, m_2)$, takes the form
\bea
&&{g}(z) = L_a \left( \begin{array}{cc} h^a_{\mu\nu}(z) & 0  \\ 0 & 0 \end{array} \right) L_a^T, \qquad  z\in I_a , \quad  a \neq H \\
&&{g}(z) = L_H \left( \begin{array}{cc} \gamma_{ij}(z) & 0  \\ 0 & 0 \end{array} \right) L_H^T, \qquad z\in I_H, 
\eea
where $\mu, \nu \in \{ 0,1 \}$, $i,j\in \{ 1,2 \}$ and $L_a$ are the transformation matrices to an adapted basis $\tilde{E}_A$ for each rod $I_a$ such that $\tilde{E}_A= (L_a^{-1})_A^{~B} E_B$. Explicitly, the adapted bases for the rods $I_L, I_H, I_D, I_R$  are the standard basis, $(m_1,m_2, k+\Omega_i m_i)$, $(k, m_1, v_D)$, $(k, m_2, m_1)$  respectively, which gives $L_L= \text{Id}$, 
\be
L_H=\left( \begin{array}{ccc} -\Omega_1 & -\Omega_2 & 1 \\ 1 & 0 & 0 \\ 0 & 1 & 0 \end{array} \right) , \qquad  L_D = \left( \begin{array}{ccc} 1 & 0 & 0 \\ 0 & 1 & 0 \\ 0 & -n & 1 \end{array} \right) , \qquad L_R = \left( \begin{array}{ccc} 1 & 0 & 0 \\ 0 & 0 & 1 \\ 0 & 1 & 0 \end{array} \right)  \; .
\ee
  The vector $\xi:= k+\Omega_i m_i$ is normal to the horizon and the constants $\Omega_i$ are the angular velocities of the black hole.
  
Next, define the Ernst potentials $b_\mu^a$ and twist potentials $\chi_i$ by
\bea
&&\td b^a_\mu = \tilde{\star} ( \tilde{E}_0\wedge \tilde{E}_1 \wedge \td \tilde{E}_\mu), \qquad a \neq H\\
&&\td \chi_i = \star (m_1 \wedge m_2 \wedge \td m_i)  \; ,
\eea
where $\tilde{\star}= \det L_a \star$ with $\star$ given by the standard orientation and define the matrices
\bea
&&X_a(z,k) := L_a \left( \begin{array}{cc} -\delta_\mu^{~\nu} & b_{\mu}^a(z) \\ 0 & 2( k-z_a) \end{array} \right) L_a^{-1}, \qquad a \neq H \\
 && X_H(z,k) := L_H \left( \begin{array}{cc} -\delta_i^{~j} & \chi_i(z) \\ 0 & 2( k-z_2) \end{array} \right) L_H^{-1}  \; ,
\eea
where $k$ is a complex (spectral) parameter and the functions $b^a_\mu(z)$ and $\chi_i(z)$ are the Ernst and twist potentials on the $z$-axis. We choose a gauge such that $b^D_\mu(z_2)=0$, $\chi_i(z_1)=0$ and $b^L_\mu, b^R_\mu\to 0 $ at infinity.  

Then, the general solution for the metric data $(h^a_{\mu\nu}(z), b^a_\mu(z))$ on each axis rod $I_a$ and $(\gamma_{ij}(z), \chi_i(z))$ on the horizon rod $I_H$ is given by the unique solution to
\be
{g}(z)= X_a(z, z) F_a(z)\;,  \qquad z\in I_a\;,   \label{gXF}
\ee
where the matrices $F_a(k)$ are given by
\bea
&&F_L(k) = -C^{-1}(P_1(k)P_2(k) P_3(k))^T, \qquad F_H(k)=- P_1(k)^{-1} C^{-1} (P_2(k)P_3(k))^T, \; \\
&&F_D(k)=- (P_1(k)P_2(k))^{-1} C^{-1} P_3(k)^T , \qquad F_R(k)= - (P_1(k)P_2(k) P_3(k))^{-1}C^{-1}  \; ,
\eea
and
\be
C:= \text{diag}(-1, 1, -1), \quad P_a(k) := X_a(z_a,k) X_{a+1}(z_a, k)^{-1} \qquad \text{for}\quad a=1, 2, 3.
\ee 
The explicit solution to (\ref{gXF}) for  $(h^a_{\mu\nu}(z), b^a_\mu(z))$  or $(\gamma_{ij}(z), \chi_i(z))$  in terms of the components of $F_a(z)$ is given in~\cite{Lucietti:2020ltw}. Here we note the important relations
\bea
&&h^a(z):=\det h^a_{\mu\nu}(z)  = - \frac{1}{\tilde{F}_{a 22}(z)} \; ,\qquad a \neq H  \label{deth} \\
&&\gamma(z):= \det \gamma_{ij}(z) =- \frac{1}{\tilde{F}_{H 00}(z)}  \; ,
\label{detgamma}
\eea
where $F_a(k):= L_a \tilde{F}_a(k) L_a^T$ defines $\tilde{F}_a(k)$ and $\tilde{F}_{a AB}(k)$ denotes the $AB$ component relative to the adapted basis for $I_a$ defined above.

The above gives the metric and Ernst/twist potentials everywhere on the axis and horizon in terms of the parameters, or moduli,  
\be
\{ \ell_H, \ell_D, b_\mu^L(z_1), \chi^H_i(z_2), b^D_\mu(z_3), b^R_\mu(z_3), \Omega_i \}  \; ,\label{moduli}
\ee
where $\ell_H := z_2-z_1, \ell_D := z_3-z_2$ are the horizon and axis rod lengths. Furthermore, these parameters must satisfy a set of nonlinear algebraic equations equivalent to
\be 
F_L(k)^T=F_L(k) \label{modspace}
\ee
for all $k$, which we refer to as the {\it moduli space equations}. In fact, this guarantees that $F_a(k)$ are symmetric matrices for all $a$.  Furthermore, these parameters must be such that 
\bea
&&h^a(z):= \det h^a_{\mu\nu}(z)  <0 \; , \qquad z\in I_a, \qquad a \neq H   \label{hlorentz}\\
&&\gamma(z):= \det \gamma_{ij}(z) >0 \; ,\qquad z\in I_H \label{gammariemann}
\eea
to ensure the metric induced on the axes and horizon are Lorentzian and Riemannian respectively.  In particular, since $\tilde{F}_{a 22}(k)$ and $\tilde{F}_{H 00}(k)$ are rational functions with simple poles at the endpoints of $I_a$ and $I_H$ respectively\footnote{This follows from the explicit solution for $F_a(k)$,  together with (\ref{deth}), (\ref{detgamma}) and the fact that $h^a(z)$ and $\gamma(z)$ must vanish at the endpoints of the associated rods.}, the signature conditions (\ref{hlorentz}) and (\ref{gammariemann}) imply
\bea
&&r^-_a:= \text{res}_{k=z_{a-1}} \tilde{F}_{a 22}(k) >0, \qquad r^+_a:= \text{res}_{k=z_{a}} \tilde{F}_{a 22}(k) <0, \qquad a \neq H  \label{resaxis} \\
&&r^-_H:=\text{res}_{k=z_{1}} \tilde{F}_{H 00}(k) <0, \qquad r^+_H:= \text{res}_{k=z_{2}} \tilde{F}_{H 00}(k) >0  \; .  \label{reshor}
\eea
This weaker form of (\ref{hlorentz}) and (\ref{gammariemann}) will be useful below.

From the solution $(h^L_{\mu\nu}, b^L_{\mu})$ on $I_L$ we can compute the mass $M$ and angular momenta $J_i$ using
\be
\begin{aligned}
&h^L_{\mu\nu}(z) = \left( \begin{array}{cc} -1 - \frac{4M}{3\pi z}+ O(z^{-2})&   \frac{2J_1}{\pi z} +O(z^{-2}) \\  \frac{2J_1}{\pi z} +O(z^{-2}) & -2z +O(1) \end{array} \right),   \\
&b^L_{\mu}(z)= \left( \begin{array}{c} -\frac{2J_2}{\pi z}+ O(z^{-2}) \\  \frac{4 \zeta}{z}+O(z^{-2}) \end{array} \right)  \; ,  
\end{aligned}
\ee
as $z\to -\infty$,  where we have fixed $b^L_\mu \to 0$ at infinity. Similarly, given the solution on $I_R$  we can compute the  asymptotic quantities from
\be
\begin{aligned}
&h^R_{\mu\nu}(z) = \left( \begin{array}{cc} -1 + \frac{4M}{3\pi z}+ O(z^{-2})&  - \frac{2J_2}{\pi z} +O(z^{-2}) \\ - \frac{2J_2}{\pi z} +O(z^{-2}) & 2z +O(1) \end{array} \right),   \\
&b^R_{\mu}(z)= \left( \begin{array}{c} -\frac{2J_1}{\pi z}+ O(z^{-2}) \\  \frac{4 \zeta}{z}+O(z^{-2}) \end{array} \right),   
\end{aligned} 
\ee
as $z\to \infty$, again fixing $b^R_\mu \to 0 $ at infinity ($\zeta$ is an asymptotic invariant which we will not consider).  

As found in~\cite{Lucietti:2020ltw} it turns out to be convenient to analyse an alternate form of the general solution above obtained by solving (\ref{gXF}) with $F_a$ replaced by $F_a^T$ for all $a \neq R$, which we assume henceforth.
Using the explicit form of the solution  we find that matching to the asymptotic expansions gives
\be
\begin{gathered}
\label{MJ}
M= \tfrac{3\pi}{4}(\ell_H+ \tfrac{1}{2} \Omega_i \chi_i(z_2)), \\
J_i= \tfrac{\pi}{4}\chi_i(z_2),     
\end{gathered}
\ee
as well as equations for $b^L_\mu(z_1), b^D_
\mu(z_3), b^R_\mu(z_3)$.\footnote{These relations can also be derived from general identities for the Ernst and twist potentials~\cite{Lucietti:2020ltw}.}
To express these it is convenient to work with dimensionless parameters
\be
\begin{gathered}
j_i := J_ i M^{-3/2} \left( \frac{27 \pi}{32} \right)^{1/2},\qquad  \omega_i := \Omega_i M^{1/2}  \left( \frac{8}{3\pi} \right)^{1/2}, \qquad \lambda_H := \frac{3\pi}{4 M} \ell_H, \\
f^D_0 := b^D_0(z_3) \left( \frac{3 \pi}{8 M} \right)^{1/2},\qquad  f^D_1 := b^D_1(z_3) \left( \frac{3 \pi}{8 M} \right), \qquad \lambda_D := \frac{3 \pi}{4 M} \ell_D \; ,
\end{gathered}
\ee
and $f^L_\mu$, $f^R_\mu$ defined similarly with $b_\mu^a(z_a)$ replaced by $b^L_\mu(z_1), b^R_\mu(z_3)$ respectively. Here and henceforth we of course assume $M>0$. Then, eliminating $\chi_i(z_2)$ in favour of $J_i$ as above, we find that the full set of asymptotic conditions are
\be
\begin{gathered}
\label{asymsBL}
    \lambda_H = 1 - \omega_i j_i,\\
      f_\mu^L = \omega_2 \begin{pmatrix}1\\-j_1\end{pmatrix} - f_\mu^D, \qquad
    f_\mu^R = \omega_1 \begin{pmatrix}-1\\j_2\end{pmatrix} + n \begin{pmatrix}f_0^D\\-\lambda_D - n f_1^D\end{pmatrix}   \;,
\end{gathered}
\ee
together with nontrivial conditions for the angular momenta
\be
\begin{aligned}
\label{jeqnsBL}
&j_1 = \omega_1 - \omega_1 j_2 ( \omega_2 - f^D_0) + (\lambda_D + n f_1^D)(\omega_1 - n f^D_0),\\
&j_2 = - f_0^D + (1- \omega_1 j_1) (\omega_2 - f^D_0) - f_1^D (\omega_1 - n f^D_0).   
\end{aligned}
\ee
The relations (\ref{asymsBL}) fix $\lambda_H, f_\mu^L, f_{\mu}^R$ in terms of the remaining parameters $\lambda_D, f_{\mu}^D, j_i, \omega_i$ which must obey the constraints (\ref{jeqnsBL}).\footnote{By combining (\ref{jeqnsBL}) one can derive the thermodynamic identity recently found using a different technique~\cite{Kunduri:2018qqt}.}

We now turn to the consistency conditions given by evaluating Ernst and twist potentials at the endpoints of their associated rods. It can be checked that $f^L_\mu(z)|_{z \to z_1} = f^L_\mu$, $f^D_\mu(z)|_{z\to z_3} = f^D_\mu$, $f^R_\mu(z)|_{z \to z_3} = f^R_\mu$ and  $\chi_i(z)|_{z\to z_2}= 4 J_i/\pi$ are all automatically satisfied (as they must be by Proposition 3 in~\cite{Lucietti:2020ltw}).  However $f^D_\mu(z)|_{z \to z_2} = 0$ gives the following constraints\footnote{Similarly, the condition $\chi_i(z)|_{z\to z_1}= \chi_i(z_1)$ gives a nontrivial constraint, although as we explain below we expect this to be redundant.}
\be
\label{fDBL}
f^D_\mu = \frac{\lambda_D}{D}
\begin{pmatrix} 
j_2 + n j_1 -( n + j_2(\omega_1 - n \omega_2))(\omega_1 - n f^D_0)  \\ 
-( n + j_2(\omega_1 - n \omega_2)) \lambda_D 
\end{pmatrix}
\ee
where
\begin{equation}
\begin{split}
    D := (1 - \omega_1 j_1)&(1 - j_1 ( \omega_1 - n \omega_2)) - (j_2 + n j_1) f^D_0\\
    &+ (2 n \lambda_D - \omega_1 j_2 + (n^2 - 1) f^D_1) ( n + j_2(\omega_1 - n \omega_2)).
\end{split}
\end{equation}
These equations are well-defined as the denominator $D\neq 0$.  This immediately follows from the relation $D=2 \lambda_H \lambda_D r_-^D$ together with (\ref{resaxis}) and positivity of the rod lengths. To see this more directly suppose $D=0$, which implies $f_\mu^D  D=0$ and hence
\begin{equation}
    j_2 + n j_1 =0, \qquad n + j_2(\omega_1 - n \omega_2) = 0 \; ,
\end{equation}
where we have used (\ref{fDBL}) and the fact that $\lambda_D>0$.
Combining these two equations and using the expression in (\ref{asymsBL}) for $\lambda_H$, one finds that $n \lambda_H = 0$ and hence if $n\neq 0$ this contradicts our assumption $\lambda_H>0$ and therefore we deduce $D\neq 0$ as claimed (one can also prove that $D\neq0$ if $n=0$~\cite{Lucietti:2020ltw}). 

To summarise, we have shown that the solution is parameterised by the mass $M>0$ and the seven dimensionless parameters $\lambda_D, f_{\mu}^D, j_i, \omega_i$ subject to the four algebraic equations (\ref{jeqnsBL}), (\ref{fDBL}) and the inequalities arising from positivity of the rod lengths
\be
\lambda_H>0, \qquad \lambda_D>0 \; ,   \label{rodlengthpos}
\ee
together with the conditions (\ref{resaxis}) and (\ref{reshor}). Generically this leaves a four-parameter family of solutions which we expect to correspond to the general unbalanced doubly spinning black lens (also a four-parameter family, proven to exist in~\cite{Khuri:2017xsc}).

More precisely, we make the following conjecture:
{\it Given a solution to (\ref{gXF}) with $F_a$ replaced with $F_a^T$ for all $a\neq R$ and the inequalities (\ref{rodlengthpos}), (\ref{resaxis}) and (\ref{reshor}), then (i) the moduli space equations (\ref{modspace}) are equivalent to (\ref{jeqnsBL}), (\ref{fDBL}) (with the rest of the moduli fixed by (\ref{MJ}), (\ref{asymsBL})), and (ii) the signature conditions (\ref{hlorentz}), (\ref{gammariemann}) are automatically satisfied.} This is a stronger form of conjecture 2 in~\cite{Lucietti:2020ltw} (which was formulated for any rod structure) specialised to the $L(n,1)$ black lens case we are considering, in the sense that we are replacing the signature conditions (\ref{hlorentz}), (\ref{gammariemann}) by their weaker form (\ref{resaxis}), (\ref{reshor}). We have been unable to prove that this conjecture holds in this case (in either form), although we find it holds for various special cases that we are able to solve analytically and we have also obtained numerical evidence for it in the generic case. In any case, we find equations (\ref{jeqnsBL}), (\ref{fDBL}) are more convenient to work with than the actual moduli space equations (\ref{modspace}), since the latter reduce to a redundant set of seven equations for $\lambda_D, f_{\mu}^D, j_i, \omega_i$ (obtained from the coefficients of $k$ in (\ref{modspace})). 

\subsection{Equilibrium conditions}
We may further restrict to regular solutions by demanding the absence of conical singularities as $\rho\to 0$ at the finite axis rod $I_D$ and also at the corner $z=z_3$ where two axis rods $I_D$ and $I_R$ meet. Firstly, the `balance condition', i.e., removal of the conical singularity as $\rho\to 0$ at $I_D$, is
\be
c_D^2:= \lim_{\rho\to 0, \,  z\in I_D} \frac{\rho^2 e^{2\nu}}{ | v_D |^2}= 1  \; .  \label{balance}
\ee
On the other hand, the metric induced on the axis component corresponding to $I_D$ is~\cite{Lucietti:2020ltw} 
\be
\mathbf{g}_D= \frac{c_D^2 \td z^2}{ | h^D(z)| }+ h^D_{\mu\nu}(z) \td x^\mu \td x^\nu  \; ,
\ee
where in these coordinates $k=\partial_{0}, m_1=\partial_{1}$. This is a smooth Lorentzian metric for $z\in I_D$ with a conical singularity at the upper endpoint $z=z_3$, which is absent iff
\be
\frac{{h^D}'(z_3)^2}{h^D_{00}(z_3)} = - 4 c_D^2  \; ,   \label{regcorner}
\ee
in which case it extends to a smooth metric on $\mathbb{R}\times D$ where $D$ is a disc topology 2-surface.  Combining this with the balance condition (\ref{balance}) gives a constraint purely on the moduli (\ref{moduli}).

Similarly, the absence of a conical singularity as $\rho \to 0$ at $I_R$ requires that we fix $c_R^2:= \lim_{\rho\to 0, \,  z\in I_R} \rho^2 e^{2\nu}/ | v_R |^2= 1$ (of course this must be obeyed for any asymptotically flat spacetime).  On the other hand, the metric induced on the axis $I_R$, which takes the same form as on $I_D$ with the obvious changes in rod labels, extends smoothly at the corner $z=z_3$ iff
\be
\frac{{h^R}'(z_3)^2}{h^R_{00}(z_3)} = - 4 c_R^2  \; .
\ee
Analogous statements hold for $I_L$.
One can derive similar relations for the metric induced on the horizon which determine the surface gravity~\cite{Lucietti:2020ltw}  (we do not display these as we will not need them).

Now, since ${g}_{00}(z_3)= h^D_{00}(z_3)=h^R_{00}(z_3)$, we can combine the balance condition (\ref{balance}) on $I_D$ and regularity condition at the corner $z=z_3$  (\ref{regcorner}), together with the analogous relations for $I_R$, to deduce the `continuity' condition\footnote{It can be shown this condition is equivalent to the continuity of $|z-z_3|  e^{2\nu}$ at $z=z_3$~\cite{Lucietti:2020ltw}.}
\be
{h^D}'(z_3)=-{h^R}'(z_3)  \; , \label{cont}
\ee
where the sign has been fixed using (\ref{hlorentz}). Using (\ref{deth}), (\ref{resaxis}) one finds ${h^D}'(z_3)=-1/r_D^+$ and ${h^R}'(z_3) = -1/r_R^-$ so that the continuity condition (\ref{cont}) is equivalent to
\be
r_D^+ + r_R^-=0  \; .\label{contalt}
\ee
More explicitly, defining $C:= -2 \lambda_D (\lambda_H+ \lambda_D)( r_D^+ + r_R^-)$, we find that (\ref{contalt}) is equivalent to
\begin{equation}
\label{conicalBL}
\begin{split}
  C=  (1 - \omega_1 j_1)&(1 - j_1 ( \omega_1 - n \omega_2) + 2\lambda_D) - (j_2 + n j_1) f^D_0\\
    &+ ((n^2 - 2) f^D_1 - \omega_1 j_2) ( n + j_2(\omega_1 - n \omega_2))\\
    &+ (n(n^2 - 2)f^D_1 + n^2(1 - \omega_2 j_2 + \lambda_D) - 1)\lambda_D = 0.
\end{split}
\end{equation}
We find that (\ref{conicalBL}) is a convenient way to impose the regularity conditions listed above, even though a priori it is only a consequence of the balance condition (\ref{balance}) and regularity at the corner $z=z_3$. Thus, by a slight abuse of terminology we will often refer to (\ref{conicalBL}) simply as the balance condition.

To summarise, we have shown that the moduli space of regular black lenses is captured by the algebraic equations (\ref{jeqnsBL}), (\ref{fDBL}) and (\ref{conicalBL}), subject to the inequalities (\ref{resaxis}), (\ref{reshor}).
Equations (\ref{jeqnsBL}), (\ref{fDBL}) and (\ref{conicalBL}) are considerably more difficult to solve than the equivalent moduli space equations for the black ring solved in~\cite{Lucietti:2020ltw} (which arises as the $n=0$ case here). For this reason it is convenient to first analyse various special cases that are tractable, such as the static limit and two distinct singly spinning limits, before considering the full doubly spinning solution.

\section{Special cases}

In this section we study two special singly spinning configurations: $J_2=0$ and $J_2+ n J_1=0$. These correspond to the Komar angular momenta with respect to $v_L$ and $v_D$ respectively, which recall are the Killing fields which possess fixed points on the horizon.  We also study the $L(n,1)$ black lens in the limit of large $n$. First though, we consider the static case.

\subsection{Static black lens \label{ssec:staticBL}}
For simplicity first suppose that $j_i=0$ for $i=1,2$. We expect this case to be static (see staticity theorem~\cite{Figueras:2009ci}) and hence necessarily singular, in line with the static uniqueness theorem~\cite{Gibbons:2002bh}.

First consider (\ref{fDBL}), which in this case reduces to
\begin{equation}
\label{fBLstatic}
    f^D_\mu = -\frac{n \lambda_D}{D}
\begin{pmatrix} 
\omega_1 - n f^D_0  \\ 
\lambda_D 
\end{pmatrix},
\end{equation}
where
\begin{equation}
\label{Dstatic}
    D = 1 + n(2 n\lambda_D + (n^2 -1) f^D_1).
\end{equation}
Eliminating $D$ between the two components of (\ref{fBLstatic}) gives $(\lambda_D+n f^D_1) (\omega_1 - nf^D_0) = \omega_1 \lambda_D$ after some algebra. Using this in the first equation of (\ref{jeqnsBL}) gives $\omega_1(1+\lambda_D) =0$ and therefore $\omega_1 = 0$ since $\lambda_D>0$. The $\mu=0$ component of (\ref{fBLstatic}) then gives that either $f^D_0=0$ or $D=n^2\lambda_D$, however the latter case can be shown to lead to a contradiction by combining it with $\mu=1$ component of (\ref{fBLstatic}) and then with (\ref{Dstatic}) and  $\lambda_D>0$. Now, the second equation of (\ref{jeqnsBL}) gives $\omega_2=0$.

Next consider the $\mu=1$ component of (\ref{fBLstatic}). To solve this it is convenient to define a new parameter $t$ according to 
\begin{equation}
\label{teqnBLS1}
    t := \frac{D}{\lambda_D} - n^2,
\end{equation}
which can be used to write $f^D_1$ as
\begin{equation}
    f^D_1 = \frac{- n \lambda_D}{t+n^2} . \label{fD1stat}
\end{equation}
Note that $t+n^2\neq0$ since, as we have already discussed, $D \neq 0$ is required by the positivity of the finite rod lengths. Using (\ref{fD1stat}) in (\ref{teqnBLS1})  with the explicit form of $D$ and solving for $\lambda_D$ in terms of $t$ gives
\begin{equation}
    \lambda_D = \frac{t + n^2}{t^2 - n^2}.
\end{equation}
This expression is well-defined since if $t^2=n^2$ that would imply that $t+n^2=0$, which as just mentioned is not allowed. Substituting this back into (\ref{fD1stat}), one can write $f^D_1$ purely in terms of $t$ to find
\begin{equation}
    f^D_1 = \frac{-n}{t^2 - n^2}.
\end{equation}

To summarise, we have shown that the asymptotic relations (\ref{asymsBL}) and the conditions (\ref{jeqnsBL}) and (\ref{fDBL}) imply $f^L_0= f^D_0 = f^R_0 = 0$ along with
\begin{equation}
    \lambda_H = 1, \qquad f^L_1= - f^D_1, \qquad f^R_1 = -n(\lambda_D + n f^D_1), \qquad \omega_i = 0  \; , 
\end{equation}
with $f^D_1$ and $\lambda_D$ determined by a single parameter $t$ as above. One can check that the moduli space equations (\ref{modspace}) are now satisfied. This is in line with our expectation that $j_i=0$ implies the solution is static.   

Finally we consider the constraints imposed by positivity of the rod lengths (\ref{rodlengthpos}) and the signature conditions (\ref{hlorentz}), (\ref{gammariemann}).  Obviously $\lambda_H>0$ is trivially satisfied, whereas $\lambda_D>0$ gives two possible branches:
\begin{equation}
\begin{split}
    \text{Branch 1:}&\qquad t > |n|\\
    \text{Branch 2:}&\qquad - n^2 < t < - |n|, \quad n\neq0,\pm1.
\end{split}
\end{equation}
As discussed in the previous section the signature conditions imply the conditions (\ref{resaxis}), which for the upper endpoint of the finite axis rod $I_D$ gives
\begin{equation}
    r^+_D = -\frac{1}{2}(t+1) < 0.  \label{rDplusstatic}
\end{equation}
This is satisfied for Branch $1$ (in which case $t>0$) and is violated for Branch $2$ (in which case $t<-2$). Thus we must discard  Branch 2. Moreover, one can show  that (\ref{hlorentz}) and  (\ref{gammariemann}) are satisfied fully for Branch $1$ demonstrating that this corresponds to the unique unbalanced static solution for this rod structure.

Finally, consider the balance condition (\ref{conicalBL}), which simplifies to
\begin{equation}
   C = \frac{t(1+t) (t^2 + n^2(1+t) )}{(t^2-n^2)^2}= 0.
\end{equation}
This condition is clearly violated for any $n \in \mathbb{Z}$ since above we have shown $t>|n|$.\footnote{In fact, for Branch 2 one can solve $C=0$ for $|n|\geq 3$. However, in this case $r_D^\pm>0$, $r_R^-<0$, which implies that the invariants $h^D$ and $h^R$ are singular at an interior point of $I_D$ and $I_R$ respectively. This is consistent with~\cite{Chen:2008fa}.} This shows that there are no regular static black lenses with this rod structure, as expected.

\subsection{A singly spinning black lens}
Next we consider the $j_2 = 0$ case of the black lens. The relations from the asymptotic conditions (\ref{asymsBL}) and (\ref{jeqnsBL}) reduce to
\be
\begin{gathered}
\label{asymsBLS1}
    \lambda_H = 1 - \omega_1 j_1,\\
    f_\mu^L = \omega_2 \begin{pmatrix}1\\-j_1\end{pmatrix} - f^D_\mu, \qquad
    f_\mu^R= \begin{pmatrix}n f^D_0 - \omega_1\\-n(\lambda_D + n f^D_1)\end{pmatrix},
\end{gathered}
\ee
and 
\begin{equation}
\begin{gathered}
\label{jeqnsBLS1}
    j_1 = \omega_1 + (\lambda_D + n f^D_1)(\omega_1 - n f^D_0),\\
    0 = - f^D_0 + (1 - \omega_1 j_1) (\omega_2 - f^D_0) - f^D_1(\omega_1 -n f^D_0), 
\end{gathered}
\end{equation}
respectively.
The consistency condition (\ref{fDBL}) becomes
\begin{equation}
\begin{split}
\label{fDBLS1}
    f^D_\mu = \frac{n \lambda_D}{D}
    \begin{pmatrix}
    j_1 - (\omega_1 - n f^D_0)\\
    -\lambda_D
    \end{pmatrix},
\end{split}
\end{equation}
where
\begin{equation}
    D = (1-\omega_1 j_1)(1-j_1(\omega_1 - n \omega_2) )   - n j_1 f^D_0 + n( 2 n \lambda_D +(n^2-1)f^D_1).
\end{equation}
We can solve these equations by introducing a new parameterisation as follows. 

Define $u$ by
\begin{equation}
    u := -\frac{j_1 - (\omega_1 - n f^D_0)}{\lambda_D},   \label{udefs1}
\end{equation}
which is well-defined since $\lambda_D>0$. This allows us to write the $\mu=0$ component of equation (\ref{fDBLS1}) as
\begin{equation}
    f^D_0 = f^D_1 u, 
\end{equation}
 and combining this with the definition of $u$ we can solve for
 \be
 j_1 = \omega_1 - u (n f^D_1 + \lambda_D).
 \ee
Substituting this into  (\ref{jeqnsBLS1}) implies\footnote{In fact (\ref{jeqnsBLS1}) admits a second solution, with $f^D_1 = -\lambda_D/n$, however this implies that $r^-_R=0$ which violates (\ref{resaxis}).}
\begin{equation}
    \omega_1 = (n f^D_1 -1)u, \qquad \omega_2 = f^D_1 u \; ,
\end{equation}
which then implies
\be
D= ((1- u^2(1+\lambda_D))^2-n^2)(1 - n f^D_1) + n^2 (1+\lambda_D).
\ee
Now define new parameters $t$ and $v$ according to
\begin{equation}
    t := \frac{D}{\lambda_D} - n^2, \qquad v := -u \sqrt{1 + \lambda_D} ,
\end{equation}
so that  just as for the static case we can write
\begin{equation}
    f^D_1 = \frac{- n \lambda_D}{t+n^2}  \; ,
\end{equation}
where $t+n^2 \neq 0$ since $D\neq  0$.
This allows one to solve the $\mu=1$ component of (\ref{fDBLS1}) for $\lambda_D$ to obtain
\begin{equation}
    \lambda_D = \frac{(t + n^2)(1 - v^2)^2}{t^2 - n^2(1-v^2)^2}.
\end{equation}
This expression is well defined since $t^2 - n^2(1-v^2)^2 = 0$ implies that $v^2=1$ which in turn implies $\lambda_H=-\lambda_D<0$ from (\ref{asymsBLS1}) which contradicts rod length positivity. Now we can express all remaining quantities in terms of $t$ and $v$:
\begin{equation}
\begin{gathered}
\label{j20Relns}
    \lambda_H =\frac{(t^2 - n^2(1-v^2))(1-v^2)}{t^2 - n^2(1-v^2)^2}, 
    \\
    j_1  = v \sqrt{1+\lambda_D}, \qquad
    \omega_i = \frac{v}{(t^2 - n^2(1-v^2)^2)\sqrt{1+\lambda_D}}
    \begin{pmatrix}
    t^2\\n(1-v^2)^2
    \end{pmatrix}  \; .
\end{gathered}
\end{equation}
Indeed, one can check that the moduli space equations (\ref{modspace}) are now satisfied without further constraint. 

The conditions $\lambda_D>0$ and $\lambda_H>0$ constrain the precise moduli space of solutions. There are three possible branches
\begin{equation}
\begin{split}
    \text{Branch 1:}& \qquad v^2<1, \quad t > \sqrt{1-v^2} |n|\\
    \text{Branch 2:}& \qquad v^2<1, \quad - n^2 < t < - \sqrt{1-v^2} |n|, \quad n\neq0\\
    \text{Branch 3:}& \qquad v^2>1+|n|, \quad -(v^2-1)|n|< t < -n^2, \quad n\neq0.
\end{split}
\end{equation}
To determine which branch actually corresponds to the unbalanced black lens consider the signature conditions (\ref{hlorentz}) and (\ref{gammariemann}). As in the static case it is helpful to consider the simpler conditions (\ref{resaxis}) on the finite axis rod $I_D$ which are
\begin{equation}
    r^-_D = \frac{1}{2\lambda_H}(t+n^2) > 0, \qquad r^+_D = -\frac{1}{2}\left(\frac{t}{1-v^2}+1\right) < 0 \; .
\end{equation}
The first condition rules out Branch $3$, whilst the second condition implies that $t/(1-v^2)>-1$, which rules out Branch $2$ (since $t/(1-v^2)<-1$ in that case). On the other hand, Branch $1$ satisfies both these conditions and furthermore it can be shown that the full conditions (\ref{hlorentz}) and (\ref{gammariemann}) are also satisfied in this case. Therefore Branch $1$  gives the unique unbalanced solution for a singly spinning black lens for this rod structure. This is as one might expect since the limit $v\to0$ of this branch gives the unbalanced static solution found in the previous section.

It is interesting to consider the moduli space of solutions in terms of physical parameters. In particular, we find that $j_1$ for the unbalanced singly spinning black lens ($n \neq 0$) (\ref{j20Relns}) is bounded above:
\be
|j_1|<  \sqrt{1+\frac{1}{|n|}} \;.    \label{j1bound}
\ee
Solutions with $j_1$ arbitrarily close to this upper bound can be found near the corner in $(t,v)$-space given by $t = |n|\sqrt{1-v^2}$ and $v = 0$. To see this, first note that  from (\ref{j20Relns}) it is easy to show that $|j_1|$ for fixed $v$  is a monotonically decreasing function of $t$ over the domain defined by Branch 1 and therefore is bounded by  its value at $t=|n|\sqrt{1-v^2} $. Then one finds that $|j_1(t=|n| \sqrt{1-v^2} , v)|$ is monotonically decreasing in $|v|$ and is thus bounded by its value at $v=0$ which is given by the RHS of (\ref{j1bound}). We remark that the curve $t = |n|\sqrt{1-v^2}$ corresponds to a component of the extremal locus $\lambda_H=0$ and therefore the upper bound on $j_1$ arises from the extremality bound. This result is in clear contrast to the singly spinning black rings ($n=0$)  for which $j_1$ can be arbitrarily large for both the unbalanced and balanced solutions (note that this is consistent with the $n\to 0$ limit of (\ref{j1bound})).

Finally,  the balance condition (\ref{conicalBL}) reduces to
\begin{equation}
  C = \frac{ t (1-v^2)^2 \left[ (n^2-1)t^2 + (t+1-v^2)(t^2 + n^2(1-v^2)) + t (1-v^2)(t+n^2)\right] }{(t^2-n^2(1-v^2)^2)^2} =0.
\end{equation}
For $n=0$ this can be solved to give $t=2v^2-1$ and $\tfrac{1}{2}<v^2<1$, which  is the regular $S^1$ spinning black ring. For $n\neq0$ the expression after the first equality is strictly positive in the domain defined by Branch 1, so the balance condition cannot be satisfied for any values of $t$ and $v$ in the moduli space of unbalanced solutions. Curiously, $C=0$ can be satisfied for $n\neq 0$ in the domain defined by Branch 2 or 3, although in either case $r_D^- r_D^+>0$ so the invariant $h^D(z)$ must change sign on $I_D$ and is thus singular at an interior point of this axis rod (this follows from (\ref{deth}) and the fact that the only possible singularities of $F_a(z)$ on $I_a$ are at its endpoints)\footnote{Similarly for Branch 2, one has $r_R^-<0$ so $h^R(z)$ must change sign on $I_R$ and thus posses a singularity on this axis.}. This proves that there are no regular singly spinning black lenses for this rod structure with $n\neq0$. This explains why the previously constructed singly spinning black lens solutions must  have conical singularities that cannot be removed or naked singularities~\cite{Chen:2008fa, Tomizawa:2019acu}. 

\subsection{A distinct singly spinning black lens}
Another natural singly spinning black lens is given by $j_2+n j_1=0$. Solving for  $j_2 = -n j_1$, the relations from the asymptotic conditions (\ref{asymsBL}), (\ref{jeqnsBL}) now give
\be
\begin{gathered}
    \lambda_H = 1 - j_1 (\omega_1 - n \omega_2)\\
    f_\mu^L = \omega_2 \begin{pmatrix}1\\-j_1\end{pmatrix} - f^D_\mu, \qquad
    f_\mu^R = -\omega_1 \begin{pmatrix}1\\n j_1\end{pmatrix} + n \begin{pmatrix}f^D_0\\-\lambda_D - n f^D_1\end{pmatrix}
\end{gathered}
\ee
and the equations
\begin{equation}
\begin{gathered}
\label{jeqnsBLS1n}
j_1 = \omega_1 + n \omega_1 j_1 ( \omega_2 - f^D_0) + (\lambda_D + n f^D_1)(\omega_1 - n f^D_0)\\
- n j_1 = - f^D_0 + (1- \omega_1 j_1) (\omega_2 - f^D_0) - f^D_1 (\omega_1 - n f^D_0).
\end{gathered}
\end{equation}
Condition (\ref{fDBL}) becomes
\begin{equation}
\label{fDBLS1n}
f^D_\mu = \frac{-n \lambda_D}{P}
\begin{pmatrix} 
\omega_1 - n f^D_0  \\ 
\lambda_D 
\end{pmatrix} ,
\end{equation}
where
\begin{equation}
\begin{split}
    P := \frac{D}{1 - j_1 ( \omega_1 - n \omega_2)} = 1 + 2n^2 \lambda_D + (n^2-1) (n f^D_1 + \omega_1 j_1).   \label{Pdef}
\end{split}
\end{equation}
Note that (\ref{fDBLS1n}) is well-defined since $P \neq 0$ as a consequence of $1 - j_1 ( \omega_1 - n \omega_2) = \lambda_H >0$ and $D\neq 0$.

\subsubsection{$L(1,1)$ horizon}
First, consider the special case $n=\pm1$.  Then the equations (\ref{jeqnsBLS1n}) and (\ref{fDBLS1n}) can be solved straightforwardly in terms of $\lambda_D$ and $\omega_1$ to give
\begin{equation}
\begin{gathered}
   j_1 = \frac{\omega_1(1+\lambda_D)}{1+\omega_1^2(1+2\lambda_D)}, \quad \omega_2 = \mp \frac{\omega_1(2(1+\lambda_D)^2-1)}{1+\lambda_D},\\   f^D_0 = \mp \frac{\omega_1 \lambda_D}{1+\lambda_D}, \quad f^D_1 = \mp \frac{\lambda_D^2}{1+2 \lambda_D}   \; ,
\end{gathered}
\end{equation}
which imply that
\begin{equation}
    \lambda_H = \frac{1- \omega_1^2(1+\lambda_D)(1+2\lambda_D)}{1+\omega_1^2(1+2\lambda_D)}.
\end{equation}
The inequalities $\lambda_H, \lambda_D >0$ are thus equivalent to the region
\begin{equation}
\lambda_D>0, \qquad    \omega_1^2<\omega_{\text{ext}}^2:= \frac{1}{(1+\lambda_D)(1+2\lambda_D)} ,  \label{msnequal1}
\end{equation}
where the upper bound for $\omega_1$ is equivalent to the extremality bound $\lambda_H=0$ (hence the notation).
The signature conditions (\ref{hlorentz}), (\ref{gammariemann}) and the moduli space equations (\ref{modspace}) now turn out to be satisfied automatically. Hence we have fully determined the moduli space in this case. 

This case also turns out to have an upper angular momentum bound given by
\be
|j_1|< \frac{1}{\sqrt{2}} \;. \label{jbound2}
\ee
This arises from the following facts: (i) $j_1(\omega_1, \lambda_D)$ is monotonic in $\omega_1$ over the domain (\ref{msnequal1}) and is thus bounded by its value at the extremality bound; (ii) $|j_1( \omega_{\text{ext}}, \lambda_D)|$ is monotonic in $\lambda_D$ and hence bounded by its value as $\lambda_D \to \infty$ which is the RHS of (\ref{jbound2}).  Observe that in this case we get a more stringent bound for $j_1$ than in the singly spinning case $j_2=0$ which from (\ref{j1bound}) gives $|j_1|<\sqrt{2}$.

Next, the balance condition (\ref{conicalBL}) in this case can be written as
\begin{equation}
C = \frac{\lambda_H (1+2 \lambda_D)}{1+\lambda_D}+ \frac{ \lambda_D( 1+ 3\lambda_D(1+\lambda_D)(2+\lambda_D))}{(1+\lambda_D)(1+2\lambda_D)(1+\omega_1^2(1+2 \lambda_D))}=0  \; ,
\end{equation}
which is clearly incompatible with $\lambda_H>0, \lambda_D>0$ and so there are no regular solutions in this class.

\subsubsection{$L(n,1)$ horizon}
We now return to the general case $n^2>1$. Since $P\neq n^2 \lambda_D$\footnote{If $P=n^2 \lambda_D$ then (\ref{fDBLS1n}) gives $\omega_1=0$ and $n f^D_1=-\lambda_D$ and (\ref{Pdef}) reduces to $\lambda_D=-1$ which violates $\lambda_D>0$.}
we can define a new nonzero parameter
\begin{equation}
\label{tBLS1n}
    t := \frac{n \lambda_D}{P - n^2 \lambda_D}.
\end{equation}
Rearranging this for $P$ gives
\begin{equation}
    P = \frac{n \lambda_D( 1+ nt)}{t}  \; ,
\end{equation}
and using this expression for $P$ in equations (\ref{fDBLS1n}) one can solve for $f^D_\mu$ to obtain
\begin{equation}
    f^D_\mu = - t
    \begin{pmatrix}
    \omega_1\\\frac{\lambda_D}{1+ n t}
    \end{pmatrix}.
\end{equation}
Note that since $P \neq 0$ we must have $1+nt\neq0$ and so the expression for $f^D_1$ is well defined. Next, substituting into the definition (\ref{Pdef}) gives an expression linear in $j_1$ which is solved to give
\begin{equation}
\label{j1BLS1n}
    j_1 = -\frac{t(1+nt)-n\lambda_D(1-t^2)}{t \omega_1 (n^2 - 1)(1+n t)} \; ,   
\end{equation}
where we assume $\omega_1 \neq 0$ ($\omega_1=0$ leads to the static black lens which we have already analysed separately in section \ref{ssec:staticBL})\footnote{To see this, note that eliminating $P$ in (\ref{fDBLS1n}) gives $(\omega_1-n f^D_0)f^D_1 = \lambda_D f^D_0$ which implies that $(\lambda_D + n f^D_1)(\omega_1 - n f^D_0)=\omega_1 \lambda_D$. Substituting this into the first equation of (\ref{jeqnsBLS1n}) gives $j_1=0$ and hence $j_2=0$. }. Now consider equations (\ref{jeqnsBLS1n}):  the first equation can be solved for $\omega_2$ and then the second can be solved for $\omega_1^2$ resulting in
\begin{eqnarray}
\label{w2BLS1n}
&&\omega_2 = \frac{t(1+nt) - n\lambda_D (1-t^2) - t \omega_1^2 ((1+nt)(1+n(t-n)(1+\lambda_D)) - (n^2-1)\lambda_D)}{n\omega_1 (t(1+nt) - n\lambda_D (1-t^2))} \; , \\
\label{w1BLS1n}
  &&  \omega_1^2 =-\frac{\lambda_D(1-t^2)(t(1+n t)- n \lambda_D(1-t^2))}{t(1+\lambda_D)(1+n t)^2(t(n+t)-\lambda_D(1-t^2))} \; ,
\end{eqnarray}
where the choice of sign for $\omega_1$ is not fixed.  It is easy to see that the denominator for $\omega_2$ is nonzero in view of our assumptions $n^2>1, \omega_1\neq 0$ and $j_1\neq 0$ together with (\ref{j1BLS1n}) (if $j_1=0$ then $j_2=0$ which is the static case).  It is also easy to show that the denominator of $\omega_1^2$ is nonzero under these assumptions\footnote{If $t(n+t)-\lambda_D(1-t^2)=0$ one can solve for  $\lambda_D=\frac{t(n+t)}{1-t^2}$ (since $n^2 \neq 1$ means that $1-t^2 \neq 0$). Then (\ref{jeqnsBLS1n}) gives $t=-n$ (since $\omega_1, j_1 \neq 0$) and hence $\lambda_D=0$, which is not allowed.} so both expressions are indeed well-defined.

Equations (\ref{jeqnsBLS1n}) and (\ref{fDBLS1n}) have now been solved in terms of $t$ and $\lambda_D$ with an additional choice of sign from $\omega_1$. Note that the moduli space equations (\ref{modspace}) are now also satisfied. In these variables $\lambda_H$ is given by
\begin{equation}
    \lambda_H = \frac{(1+n t)(t^2(n+t)(1+ \lambda_D) - n (1- t^2)\lambda_D^2)}{t \lambda_D(n^2-1)(1-t^2)}.
\end{equation}
The moduli space is defined by $\lambda_H, \lambda_D, \omega_1^2 > 0$ along with the conditions (\ref{hlorentz}), (\ref{gammariemann}). As for the special cases considered previously it is convenient to  consider these latter conditions in the weaker form (\ref{resaxis}) and in particular 
\begin{equation}
    r_D^+ = -\frac{n+t}{2t}<0.
\end{equation}
By combining this  with $\lambda_H, \lambda_D, \omega_1^2 > 0$, one can show that $t$ and $\lambda_D$ must satisfy
\begin{equation}\label{mssspin2}
\begin{gathered}
    \frac{t(1+n t)}{n(1-t^2)}<\lambda_D<\lambda_D^{\text{ext}}:= \frac{t^2(n+t)+t\sqrt{t^2(n+t)^2+4n(1-t^2)(t+n))}}{2 n (1-t^2)} \; ,\\
    n t>0 ,\qquad 1-t^2>0.
\end{gathered}
\end{equation}
These conditions turn out to be sufficient to imply the full conditions (\ref{hlorentz}) and (\ref{gammariemann}) and so describe the full space of unbalanced solutions in this case. The upper bound on $\lambda_D$ corresponds to extremal solutions with $\lambda_H=0$ and the lower bound corresponds to the static limit of Section \ref{ssec:staticBL} (though note that a different parameter $t$ is used there).

One  again finds a  bound on the angular momentum which in this case is given by,
\begin{equation}
     |j_1| < \sqrt{\frac{1}{|n|(1+|n|)}} \; .  \label{jbound2gen}
\end{equation}
This follows from the fact that (\ref{j1BLS1n}) for fixed $t$ is monotonically increasing in $\lambda_D$ in the region (\ref{mssspin2}),  so it is bounded by its value at $\lambda_D=\lambda_D^{\text{ext}}$. Next one can show that $| j_1(t, \lambda_D^{\text{ext}})|$ is monotonically increasing in $|t|$ so it is bounded by its value at $|t|=1$ which gives (\ref{jbound2gen}). Note that (\ref{jbound2gen}) agrees with the $n=\pm 1$  special case we found above (\ref{jbound2}). Thus for general $n$ we again have a more stringent bound for $j_1$ than in the $j_2=0$ singly spinning case (\ref{j1bound}).

Finally consider the balance condition (\ref{conicalBL}). This can be written as
\begin{equation}
\begin{aligned}
C = \frac{\lambda_D}{n^2-1} + \frac{ \lambda_H \lambda_D (n+2 t)}{t(1+n t)}+ \frac{t(n+t)((n^2-2)+nt)(1+\lambda_D)}{(1-t^2)(n^2-1)}= 0  \; ,
\end{aligned}
\end{equation}
which cannot be satisfied for any $t$ and $\lambda_D$ in the moduli space (\ref{mssspin2}) since each of the three terms after the first equality is manifestly positive. This proves that for $n^2>1$ there are no regular singly spinning solutions of this kind.

\subsection{Myers-Perry limit}
\label{sec:largen}
Next we consider the question of what happens to these $L(n,1)$ black lens solutions as $n$ becomes arbitrarily large.  The rod vector on the finite axis rod $v_D=(n,1)$ clearly diverges in this limit, however, since any multiple of the rod vector also vanishes on $I_D$ we can rescale it so that it has a finite limit to obtain $n^{-1} v_D\to (1,0)= v_R$ as $n \to \infty$ (of course this will spoil the periodicity of $v_D$ for finite $n$, although not in the limit $n \to \infty$). Therefore,  in this limit $I_D$ becomes part of $I_R$ and hence the rod structure as $n \to \infty$ reduces to that of the Myers-Perry black hole. This suggests that the $n \to \infty$ limit of the unbalanced $L(n,1)$ black lens solution should be the Myers-Perry solution.  This turns out to be the case and can be seen to emerge from our moduli space equations as follows.

Recall the moduli space of unbalanced solutions is given by  four equations (\ref{jeqnsBL}) and (\ref{fDBL}) for seven parameters $f^D_\mu, j_i, \omega_i, \lambda_D$.\footnote{For simplicity, in this section we are assuming the validity of the conjecture stated at the end of Section \ref{sec:gensol}.} We take $\omega_i, \lambda_D$ as the independent parameters and solve for $f^D_\mu, j_i$ by assuming they are finite (possibly vanishing) in the $n\to \infty$ limit and that they admit an expansion in $n^{-1}$. Then, expanding (\ref{fDBL})  in $n^{-1}$ we find that the leading order terms immediately give  $f^D_0 = O(n^{-1})$ and $f^D_1 = -\lambda_D/n + O(n^{-2})$. Equations (\ref{jeqnsBL}) and (\ref{asymsBL}) then imply that 
\begin{equation}
\begin{gathered}
    j_1=\frac{{\omega}_1(1-{\omega}_2^2)}{1-{\omega}_1^2 {\omega}_2^2} + O(n^{-1})\; , \qquad  j_2=\frac{{\omega}_2(1-{\omega}_1^2)}{1-{\omega}_1^2 {\omega}_2^2} + O(n^{-1})\\
    \lambda_H= \frac{(1-\omega_1^2)(1-\omega_2^2)}{1-\omega_1^2\omega_2^2} + O(n^{-1}).
\end{gathered}
\end{equation}
The leading order terms here are precisely the expressions for the Myers-Perry solution~\cite{Lucietti:2020ltw}. Next we consider the bounds on the moduli for these unbalanced solutions. In particular from (\ref{resaxis}) we obtain
\begin{equation}
    r_H^- = -\frac{3 \pi}{16 M} (1-\omega_1^2) + O(n^{-1}) <0, \qquad r_H^+ = \frac{3 \pi}{16 M} (1-\omega_2^2) + O(n^{-1}) > 0  \; ,
\end{equation}
which immediately gives $|\omega_i|<1$ to leading order and is sufficient to imply the remaining inequalities that define the moduli space $\lambda_H>0$, (\ref{hlorentz}) and (\ref{gammariemann}). These are also the moduli space conditions for Myers-Perry solutions and in particular imply that $|j_1|+|j_2|<1$ to leading order.

Now we consider the balance condition (\ref{conicalBL}) in this limit.  Using the expansions of $f^D_\mu$ and $j_i$ found above, the leading term in the large $n$ expansion of (\ref{conicalBL}) fixes the $n^{-2}$ term in $f^D_1$:
\begin{equation}
    f^D_1 = -\frac{\lambda_D}{n}+ \frac{\omega_1 \omega_2 \lambda_D (1-\omega_1^2)}{n^2((1-\omega_2^2) + \lambda_D (1-\omega_1^2\omega_2^2))} + O(n^{-3}).
\end{equation}
Note that the conditions $|\omega_i|<1$ ensure the denominator is nonzero. Next, one can solve the $\mu=0$ component of (\ref{fDBL}) for the $n^{-1}$ term in the expansion of $f^D_0$ provided $\omega_1\omega_2\neq 0$ (this involves the $n^{-1}$ terms in the expansion of $j_1$ and $j_2$). Finally, using these expansions the first equation of (\ref{jeqnsBL}) becomes
\begin{equation}
   \frac{\omega_2 (1-\omega_1^2) (\lambda_D + \lambda_H)}{n((1-\omega_2^2) + \lambda_D (1-\omega_1^2\omega_2^2))} + O(n^{-2}) = 0
\end{equation}
for $\omega_1\omega_2 \neq 0$.
The coefficient of the leading term is nonzero since $|\omega_i|<1, \lambda_D, \lambda_H>0$, resulting in a contradiction for large enough $n$. For $\omega_1 \omega_2=0$ one has to repeat the above analysis, although once again one finds that no regular solutions are possible in this limit. This demonstrates that no regular solutions exist in the large $n$ limit.

It is worth noting that there is another, more obvious, limit of the unbalanced black lens which reduces to the Myers-Perry solution. This is $\lambda_D\to 0$ with all other parameters held fixed, which corresponds to simply shrinking the finite axis rod $I_D$ away, resulting in the same rod structure as the Myers-Perry solution.  One can see this explicitly  by solving our moduli space equations in this limit, which  imply 
\be
f_0^D= \omega_2 \lambda_D + O(\lambda_D^2),  \qquad f_1^D= O(\lambda_D^2)
\ee 
and $j_i, \lambda_H, r_H^\pm$ are given by the above expressions with $O(n^{-1})$ replaced by $O(\lambda_D)$. Thus to leading order $|\omega_i|<1$ and the $j_i$ are given by the Myers-Perry expressions. Then, the balance condition (\ref{conicalBL}) in this limit is
\be
C=\frac{(1-\omega_1^2)^2}{1-\omega_1^2\omega_2^2}+ O(\lambda_D) = 0  \; ,
\ee
which clearly cannot be satisfied for small enough $\lambda_D$ since the first term is positive. Thus no regular solutions exist in this limit either.

\section{Doubly spinning black lens}

We now study the general moduli space equations for the black lens (\ref{modspace}) together with the balance condition (\ref{conicalBL}). On general grounds one would expect the moduli space of regular solutions to fill out a $2$-dimensional subset of the $(j_1, j_2)$ plane (or be empty). To see this, first note that from an existence theorem we know there must be a $4$-dimensional moduli space of unbalanced black lens solutions parameterised by $(J_1, J_2)\in \mathbb{R}^2$ and $\ell_H>0, \ell_D >0$~\cite{Khuri:2017xsc}. Therefore, since we expect the balance condition on the finite axis rod to reduce the number of parameters by one, it is reasonable to expect that the moduli space of regular solutions is $3$-dimensional (if it is non-empty). For a given mass $M>0$ this is then equivalent to a $2$-dimensional subset of the $(j_1, j_2)$-plane.  For reference, for the Myers-Perry $S^3$ black holes this region is simply $|j_1|+|j_2|<1$, whereas for the black ring it is an unbounded region~\cite{Emparan:2008eg} (we plot this in our figures below).

The moduli space equations (\ref{modspace}) and balance condition (\ref{conicalBL}) possess the following discrete symmetries:\footnote{It is easy to see this is the case for (\ref{jeqnsBL}), (\ref{fDBL}).}
\bea
&&g_1: ( n, j_1, \omega_1, j_2, \omega_2, f_0^D,f_1^D, \lambda_D)\to ( -n,- j_1,- \omega_1, j_2, \omega_2, f_0^D,- f_1^D, \lambda_D),   \\
&&g_2: ( n, j_1, \omega_1, j_2, \omega_2, f_0^D,f_1^D, \lambda_D)\to ( -n,j_1, \omega_1, - j_2, -\omega_2,- f_0^D,- f_1^D, \lambda_D).
\eea
The $g_1$-symmetry originates from the orientation-reversing symmetry $m_1\to -m_1, n\to -n, v_R\to - v_R$, whereas the $g_2$-symmetry arises from $m_2\to -m_2, n \to -n, v_D\to -v_D, v_L\to - v_L$.
Using these we may restrict to the region $j_1 \geq 0$ and $n> 0$, which we will henceforth (we exclude the $n=0$ case as that corresponds to the black ring).

Unfortunately, we have been unable to solve the moduli space equations (\ref{modspace}) (or indeed equations (\ref{jeqnsBL}), (\ref{fDBL}))  and balance condition (\ref{conicalBL}) analytically in general. Nevertheless, we have verified numerically that for a large sample of  solutions to these equations at least one of the inequalities $\lambda_H>0, \lambda_D>0$,  (\ref{resaxis}), (\ref{reshor}) is violated. We give more details on our numerical checks below. 

\begin{figure}[h!]
\begin{centering}
\includegraphics[width=16cm]{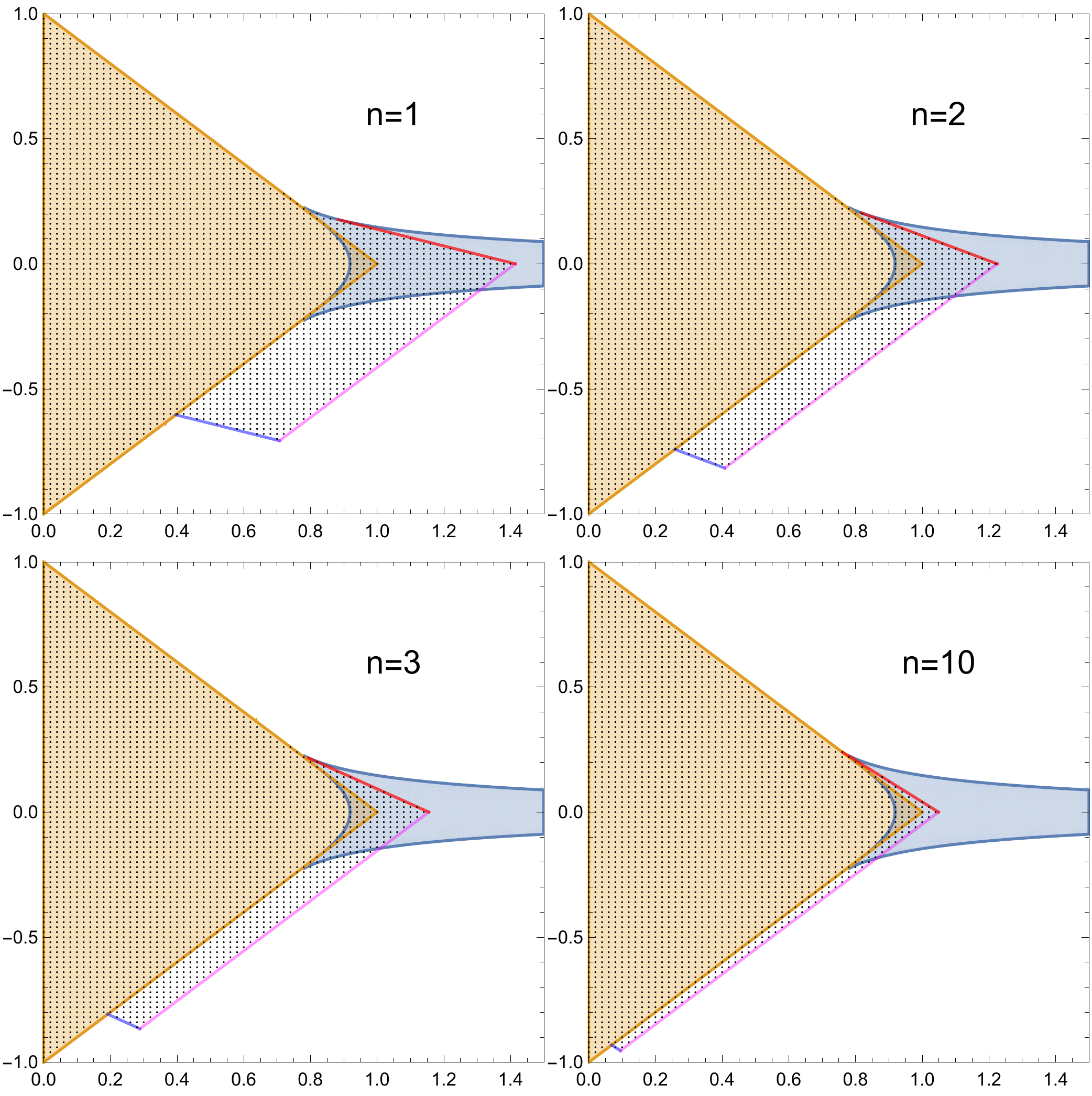}
\caption{The unbalanced $L(n,1)$ black lens solution projected to the $(j_1,j_2)$ plane (black dots). The shaded regions correspond to the Myers-Perry black hole (orange) and balanced black ring (blue). The line segments for $j_2<0$ are (\ref{lowerj2}) (blue, pink) and  for $j_2>0$ is (\ref{upperj2}) (red).}
\label{fig:lownphasespace}
\end{centering}
\end{figure}

\begin{figure}[h!]
\centering
\includegraphics[width=14cm]{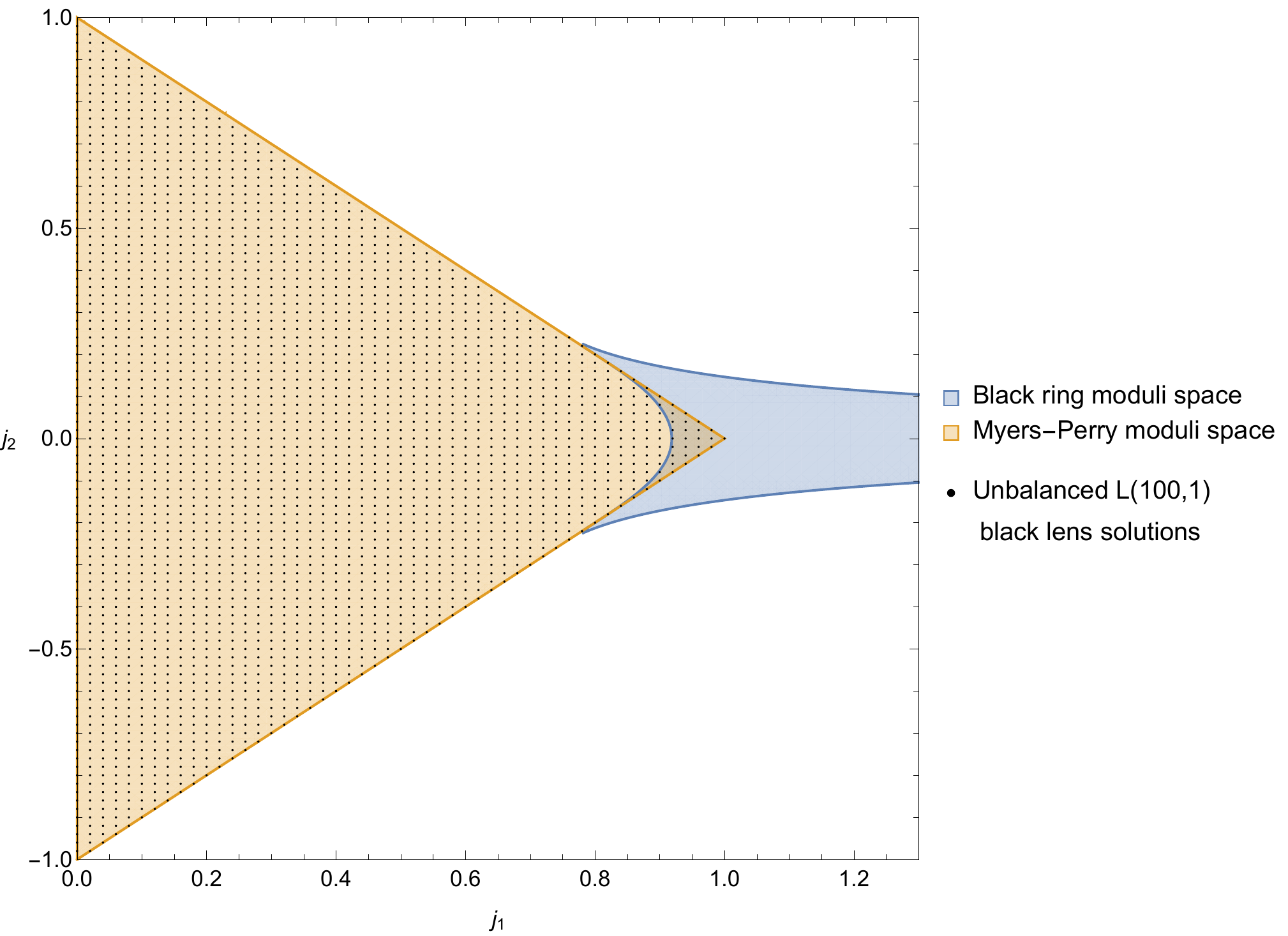}
\caption{The unbalanced $L(100,1)$ black lens solution projected to the $(j_1,j_2)$ plane. This closely approximates the Myers-Perry moduli space, consistent with the analysis in Section \ref{sec:largen}.}
\label{fig:n100PhaseSpace}
\end{figure}

First, it is instructive to consider the moduli space of solutions to (\ref{modspace}) without imposing the balance condition. As explained above this should correspond to a 4-dimensional space, or, in terms of our dimensionless variables, a 3-dimensional space.  We have numerically solved these equations and plotted the projection of this space to the $(j_1, j_2)$ plane for $n=1,2,3,10$ in Figure \ref{fig:lownphasespace} and $n=100$ in Figure \ref{fig:n100PhaseSpace}. Specifically, these plots were obtained by numerically solving the moduli space equations (\ref{modspace}) at values of $j_1$ and $j_2$ centred around the origin (on a square grid of spacing $0.02$) and values of $\lambda_D$ from $0$ to $10^3$ with greatest density of sampling in the interval $0<\lambda_D<1$. Then the plotted points correspond to solutions of these equations which also satisfy $\lambda_H>0$, (\ref{hlorentz}) and (\ref{gammariemann}).

These plots suggest that the moduli space for  unbalanced $L(n,1)$ black lenses is a bounded region in the $(j_1,j_2)$-plane somewhat akin to that of the Myers-Perry solution.  As $n$ increases this region approaches that of the Myers-Perry solution, indeed, the $n=100$ solution already closely approximates the Myers-Perry moduli space, in line with our large $n$ analysis in Section \ref{sec:largen}. For general $n$, the upper bound for $j_1$ is clearly determined by that of the $j_2=0$ solution and is consistent with the analytic bound (\ref{j1bound}) we found in this special case.  The bounds for $j_2$ are clearly not symmetric about the $j_1$-axis and for $j_2<0$ they are consistent with the analytic bound we found for the $j_2=-n j_1$ case (\ref{jbound2gen}).    Note that the lack of symmetry about the $j_1$-axis is a consequence of the rod structure and that we have fixed the discrete $g_2$-symmetry discussed above. 

In more detail, the plots  suggest that the boundary of the moduli space consists of several  segments.   For $j_2<0$ there appear to be three boundary segments which, based on the numerics and the analytic upper bounds (\ref{j1bound}), (\ref{jbound2gen}), we conjecture are:
\be
j_2= \left\{  \begin{array}{c}  j_1-1\qquad  \qquad \qquad 0<j_1< q_n \\  \frac{n}{n+2} \left(- j_1 - s_n \right) \qquad q_n<j_1< r_n  \\   j_1-s_n \qquad \qquad  \qquad r_n  <j_1<s_n  \end{array} \right. \; ,  \label{lowerj2}
\ee
where $s_n, r_n$ denotes the $j_1$ upper bound in (\ref{j1bound}), (\ref{jbound2gen}) respectively, and
\be
q_n:= \frac{1- \tfrac{1}{2}n(s_n-1)}{n+1}  \; .
\ee
In particular, for $n=1,2,3,10$ we find $q_n \approx 0.396, 0.258, 0.192, 0.069$  with $q_n\to 0$ as $n\to \infty$, which are consistent with the numerics.
The first line segment coincides with part of the Myers-Perry boundary; the second line segment  starts  at the Myers-Perry boundary and ends at the upper limit of the $j_2=-n j_1$ solution (\ref{jbound2gen}) (blue line in the plots); the third line segment ends at the upper limit of the $j_2=0$ solution (\ref{j1bound}) (pink line in the plots). 

For $j_2>0$ the numerics indicate that there are also three boundary segments, although from the plots displayed here only two are apparent. The first is a line segment given by  part of the Myers-Perry boundary $j_2=1-j_1$  for $0<j_1<u_n<3/4$; the third line segment is\footnote{The blue and red lines are parallel, intersect the $j_2$ axis at equal and opposite values of $j_2$, and the red line ends at $(s_n,0)$ and the blue line ends at  $(r_n, -n r_n )$. In fact, one can check that the blue, red and pink lines correspond to extremal $\lambda_H=0$, $\lambda_D\to \infty$,  solutions to the moduli space equations.} (red line on the plots)
\be
j_2= \frac{n}{n+2} (- j_1 + s_n) \; , \qquad  \frac{3}{4}<v_n  <j_1<s_n \; . \footnote{The value of $v_n$ seems to be close to the intersection of the upper curve for the black ring moduli space with the line defined in (\ref{upperj2}). This is given by $v_n = \frac{2+3n}{4\sqrt{n+n^2}}$ and is what has been plotted for the endpoint of that line.} \label{upperj2}
\ee
The second segment is a curve which joins these two lines although it is not visible on the plots (we will not study this here).  Note that $j_1= 3/4$ corresponds to where the balanced black ring and the Myers-Perry  moduli space boundaries meet.

It is interesting to note that for both the $j_2=0$ and $j_2=-n j_1$ singly spinning special cases that we studied analytically one has the bound
\be
|j_1|+ |j_2| < \sqrt{1+\frac{1}{|n|}} \; . \label{universalbound}
\ee
It is tempting to conjecture that this bound is satisfied for all doubly spinning unbalanced black lenses (although the moduli space does not fill out the whole of this region as can be seen from the plots). Indeed inspecting the plots, this inequality seems to be saturated for the part of the moduli space boundary between $j_2=0$ and $j_2=-n j_1$ illustrated by the pink line segments in Figure \ref{fig:lownphasespace}, which correspond to extremal $\lambda_H=0$, $\lambda_D \to \infty$ solutions. 

We now return to the question of regular black lenses, i.e., solutions in the moduli space of unbalanced black lenses which also obey the balance condition (\ref{conicalBL}).  For all the points sampled in Figures \ref{fig:lownphasespace} and \ref{fig:n100PhaseSpace} we find that $C$ defined in (\ref{conicalBL}) is positive.  We have also performed further searches by numerically solving  (\ref{modspace}) and (\ref{conicalBL}) for $n=1,2,3,10,100$, sampling $(j_1,j_2)$ in a square grid in the region $-1\le j_2\le1,0\leq j_1 \leq 1.5$ with a spacing of $0.015$. For all points we find  at least one of the inequalities $\lambda_H>0, \lambda_D>0$,  (\ref{resaxis}), (\ref{reshor}) is violated. We therefore conclude that regular black lenses do not exist in this class.

\section{Discussion}
\label{sec:disc}

We have presented evidence that regular, asymptotically flat, stationary, vacuum black holes with lens space $L(n,1)$ topology and two commuting axial Killing fields do not  exist, for the simplest possible rod structure (see Figure \ref{fig:BL}). Our evidence is based on an analytic proof that the conical singularities on the inner axis rod cannot be removed for two different singly spinning cases ($J_2=0$ and $J_2+n J_1=0$), together with numerical evidence for the generic doubly spinning case. In particular, based on the examples studied in this paper, we conjecture that any solution to the black lens moduli space equations  (\ref{modspace}) (or  (\ref{jeqnsBL}), (\ref{fDBL})) will give $C>0$ and hence violate the balance condition (\ref{conicalBL}). Of course, it would be desirable to provide a fully analytic proof of this, ideally by solving the moduli space equations (\ref{modspace}) together with the balance (regularity) condition (\ref{conicalBL}). We expect that solving the moduli space equations reduces to solving equations (\ref{jeqnsBL}), (\ref{fDBL}) as we have found in other cases (thus establishing the validity of the conjecture stated at the end of Section \ref{sec:gensol}).  Perhaps this could be achieved by finding an alternate parameterisation of the moduli space.  

The general solution for the black lens on the axes and horizons which we have analysed was constructed in previous work~\cite{Lucietti:2020ltw}. In fact, in that work the general solution on the axis and horizon for any rod structure was obtained together with the corresponding moduli space equations. In principle the rod structure can be arbitrarily complicated, since even for single black hole solutions one may have an arbitrary number of finite axis rods. These finite axis rods correspond to noncontractible 2-cycles in the domain of outer communication (possibly ending on the horizon). A natural question is whether such exotic rod structures must always lead to conically singular vacuum solutions. Unfortunately, given the difficultly in analysing the moduli space equations for the black lens rod structure in this paper,  it seems difficult to answer the general question definitively without further insights into the structure of these equations.

It is interesting to compare our results to analogous supersymmetric solutions in five-dimensional minimal supergravity. In that case, regular $L(n,1)$ black lenses and $S^3$-black holes with 2-cycles in the DOC are known~\cite{Kunduri:2014iga, Kunduri:2014kja, Tomizawa:2016kjh, Horowitz:2017fyg}, demonstrating that at least in the presence of supersymmetry, nontrivial rod structures can be realised.  In fact, a complete classification of asymptotically flat, supersymmetric and biaxisymmetric black hole and soliton solutions has been obtained, revealing an infinite class of new black hole/ring/lens solutions with 2-cycles in the DOC~\cite{Breunholder:2017ubu, Breunholder:2018roc}.  A natural physical explanation for the existence of such configurations is that the presence of a Maxwell field allows magnetic flux  to `support' the 2-cycles.  Indeed, even in the absence of a black hole, this theory possesses supersymmetric soliton solutions, i.e., asymptotically flat, stationary, everywhere regular spacetimes with positive energy, which posses 2-cycles supported by magnetic flux~\cite{Bena:2005va}. Furthermore, some of the aforementioned supersymmetric $S^3$-black hole spacetimes with nontrivial topology can be interpreted as black holes sitting in such soliton spacetimes~\cite{Horowitz:2017fyg, Breunholder:2018roc}.

The supersymmetric classification reveals that the only regular supersymmetric solutions with the rod structure studied in the present paper (see Figure \ref{fig:BL}) is the original $L(2,1)$ black lens for $|n|=2$~\cite{Kunduri:2014kja} and the black ring for $n=0$~\cite{Elvang:2004rt}  (the supersymmetric $L(n,1)$ black lenses for $|n|>2$ necessarily posses extra finite axis rods~\cite{Tomizawa:2016kjh, Breunholder:2017ubu}).  Therefore, at least for $|n|=2$, one expects that regular near-supersymmetric  black lenses should exist.\footnote{Existence of soliton and non-extremal black hole solutions, with potential conical singularities on the inner axis rods, has been recently proven in this theory~\cite{Alaee:2019qhj}.}   In contrast, here we have found that regular $|n|=2$ vacuum black lenses do not exist.  A simple physical interpretation of this is that rotation alone is not sufficient to support nontrivial topology, although the presence of a sufficiently strong magnetic flux is. Indeed, this is consistent with the nonexistence of vacuum soliton spacetimes. Perhaps this suggests that the same goes for more complicated rod structures with a single horizon.  If so then the black ring would be an exceptional case for which rotation alone is sufficient to support  nontrivial topology.  \\

\noindent {\bf Acknowledgements}. FT is supported by an EPSRC studentship. JL  is supported by  a Leverhulme Trust Research Project Grant.

\end{document}